# Folding membrane proteins by deep transfer learning

Sheng Wang[1,2,4,*], Zhen Li[1,3,*], Yizhou Yu[3], Jinbo Xu[1,†]

[1]Toyota Technological Institute at Chicago, IL USA 60637; [2]Department of Human Genetics, University of Chicago, IL USA 60637; [3]Department of Computer Science, University of Hong Kong, Hong Kong; [4]Computational Bioscience Research Center (CBRC), King Abdullah University of Science and Technology (KAUST), Thuwal, Saudi Arabia

* These authors contributed equally.

† Lead contact: Dr. Jinbo Xu, jinboxu@gmail.com

## Summary

Computational elucidation of membrane protein (MP) structures is challenging partially due to lack of sufficient solved structures for homology modeling. Here we describe a high-throughput deep transfer learning method that first predicts MP contacts by learning from non-membrane proteins (non-MPs) and then predicting three-dimensional structure models using the predicted contacts as distance restraints. Tested on 510 non-redundant MPs, our method has contact prediction accuracy at least 0.18 better than existing methods, predicts correct folds for 218 MPs (TMscore>0.6), and generates three-dimensional models with RMSD less than 4Å and 5Å for 57 and 108 MPs, respectively. A rigorous blind test in the continuous automated model evaluation (CAMEO) project shows that our method predicted high-resolution three-dimensional models for two recent test MPs of 210 residues with RMSD ~2Å. We estimated that our method could predict correct folds for 1,345–1,871 reviewed human multi-pass MPs including a few hundred new folds, which shall facilitate the discovery of drugs targeting at membrane proteins.

## Introduction

Membrane proteins (MPs) are important for drug design and have been targeted by approximately half of current therapeutic drugs (Yıldırım et al., 2007). MPs are also important in cell-environment interactions by serving as environment sensing receptors, transporters, and channels. It is estimated that 20-40% of all genes in most genomes encode MPs (Uhlén et al., 2015, Krogh et al., 2001) and larger genomes contain a higher fraction of MPs (Wallin and Heijne, 1998). For example, the human genome has >5,000 reviewed MPs, of which 3000 are non-redundant at 25% sequence identity level. Experimental determination of MP structures is challenging as they are often too large for nuclear magnetic resonance (NMR) experiments and difficult to crystallize (Lacapere et al., 2007). Currently there are only about 510 non-redundant MPs with solved structures in PDB (Kozma et al., 2012) (Supplementary Table 1), so it is important to develop computational methods to predict MP structures from sequence information.

Although knowledge-based computational prediction works well for many soluble proteins, it faces some challenges for MPs partially due to lack of sufficient MPs with solved structures. First, homology modeling only works on a small portion of MPs and thus, de novo prediction or ab initio folding is needed. Second, accurate estimation of the parameters of a computational method for MP structure prediction rely on sufficient statistics, which sometimes cannot be fulfilled due to a small number of non-redundant MPs with solved structures. So far the most successful methods for ab initio folding are based upon fragment assembly implemented in Rosetta(Kim et al., 2004) and others, but they work mostly on some small proteins. Recently, contact-assisted ab initio folding has made exciting progress. It predicts the contacts of a protein in question and then uses predicted contacts as distance restraints to guide ab initio folding. However, contact-assisted folding heavily depends on accurate prediction of protein contacts, which cannot be achieved by pure co-evolution methods on proteins without many sequence homologs.

Here we define that two residues form a contact if the Euclidean distance of their $C_\beta$ atoms is less than 8Å. Evolutionary coupling analysis (ECA) and supervised learning are two popular contact prediction methods. ECA predicts contacts by identifying co-evolved residues, such as EVfold (Marks et al., 2011), PSICOV (Jones et al., 2012), CCMpred (Seemayer et al., 2014), Gremlin (Kamisetty et al., 2013) and others(Weigt et al., 2009), but it needs a large number of sequence homologs to be effective (Skwark et al., 2014, Ma et al., 2015). Supervised learning predicts contacts from a variety of information, e.g., SVMSEQ(Wu and Zhang, 2008), CMAPpro (Di Lena et al., 2012), PconsC2 (Skwark et al., 2014) and PconsC3 (Skwark et al., 2016), MetaPSICOV (Jones et al., 2015), PhyCMAP (Wang and Xu, 2013) and CoinDCA-NN (Ma et al., 2015). CCMpred has similar accuracy as Gremlin and plmDCA, but on average better than PSICOV and Evfold (maximum-entropy option). MetaPSICOV used two cascaded 1-layer neural networks and performed the best in CASP11. CMAPpro (Di Lena et al., 2012) uses a deep architecture, but its performance was worse than MetaPSICOV in CASP11 (Monastyrskyy et al., 2015). There are some contact prediction methods specifically developed for MPs. They employ some MP-specific features and are trained from a limited number of MPs, such as TMHcon (Fuchs et al., 2009), MEMPACK (Nugent and Jones, 2010), TMhit (Lo et al., 2009), TMhhcp (Wang et al., 2011), MemBrain (Yang et al., 2013), COMSAT (Zhang et al., 2016a) and OMPcontact (Zhang et al., 2016b). McAllister and Floudas (McAllister and Floudas, 2008) proposed a mixed integer programming method for MP contact prediction by optimizing an energy function subject to a set of physical constraints. Although so many methods have been developed, for MPs with fewer sequence homologs, the predicted contacts by existing methods are still of low quality (Adhikari et al., 2015).

Previously, we developed a deep learning method mainly for soluble protein contact prediction(Wang et al., 2017), which obtained the highest F1 score in CASP12. Here we systematically study deep learning methods for MP contact prediction and folding. We predict MP contacts by concatenating two deep residual neural networks (He et al., 2015), which have won the ILSVRC and COCO 2015 competitions on image recognition (Russakovsky et al., 2015, Lin et al., 2014), and then predict 3D models by feeding predicted contacts to CNS (Brunger, 2007). Different from soluble protein contact prediction, the challenge of applying deep learning to MP contact prediction is lack of sufficient training data since there are only 510 non-redundant MPs with solved structures. To overcome this, we train our deep learning model using thousands of non-MPs with solved structures. That is, we transfer knowledge learned from non-MPs to MP protein contact prediction and thus, call such a method deep transfer learning. The resultant deep model works well for MP contact prediction, outperforming that trained by only MPs and existing methods. Our further study indicates that using a mix of non-MPs and MPs, we can train a deep model with even better performance, especially in transmembrane regions. Experimental results show that with our predicted contacts we can correctly fold 218 and 288 of 510 MPs when TMscore≥0.6 and ≥0.5 are used as the threshold, respectively, and that we can fold 57 and 169 MPs with RMSD<4.0Å and <6.5Å from their native structures, respectively. We estimate that our method can fold 1871 and 1345 of 2215 reviewed human multi-pass membrane proteins (of <700 residues) with TMscore ≥ 0.5 and 0.6, respectively. A rigorous blind test in CAMEO shows that our method produced high-resolution 3D models for two test MPs of 212 residues and correct folds for the other two.

## Results

### Deep learning model architecture

We designed a deep learning model that uses two residual neural networks (ResNet) (Figure 1). Each ResNet is composed of some blocks, each block in turn consisting of 2 batch normalization layers(Ioffe and Szegedy, 2015), 2 ReLU activation layers (Nair and Hinton, 2010) and 2 convolution layers. The first ResNet conducts 1-dimensional (1D) convolutional transformations of sequential features to capture sequential context of a residue. Its output is converted to a 2-dimensional (2D) matrix by an operation called outer concatenation (similar to outer product) and then fed into the 2[nd] ResNet together with existing pairwise features. The 2[nd] ResNet conducts 2D convolutional transformations of its input to capture the 2D context information of a residue pair (mostly contact occurrence patterns). Finally, the output of the 2nd network is fed into logistic regression to predict contact probability of any two residues. The filter size (i.e., window size) used by a 1D convolution layer is 15 while that used by a 2D convolution layer is 5×5.

Our deep learning method is unique in several respects. First, it uses a concatenation of two deep ResNets, which to our knowledge has not been applied to MP contact prediction before. Second, we predict all contacts of an MP simultaneously, as opposed to existing supervised methods that predict contacts individually. This facilitates the learning of contact occurrence patterns.

We train our deep learning model with three strategies: MP only (i.e., only MPs are used as the training set), Non-MP only (i.e., only non-MPs are used as the training set) and Mixed (both non-MPs and MPs are used as training set). See STAR Methods for details.

**Contact prediction accuracy**

All our three model training strategies outperform the other methods in terms of medium- and long-range prediction accuracy (Table 1). Notably, our deep model trained by only non-MPs outperforms our model trained by only MPs even if only the predicted contacts in transmembrane regions are evaluated. This result suggests that non-MPs and MPs share some common properties in contacts that can be learned by our deep learning model, and that the set of non-MPs contain more information for contact prediction than the set of MPs. Our deep learning method obtains the best performance especially for contacts in transmembrane regions when both non-MPs and MPs are used in the training set. MetaPSICOV is also mainly trained by non-MPs, but it performs worse than our NonMP strategy.

**Table 1.** Overall contact prediction accuracy on 510 membrane proteins. 'MP', 'NonMP', and 'Mixed' represent our deep models trained by MPs only, by non-MPs only and by a mix of MPs and non-MPs.

| Contacts in all regions are evaluated | | | | | | | | | | | | |
|---|---|---|---|---|---|---|---|---|---|---|---|---|
| Method | Short | | | | Medium | | | | Long | | | |
| | L/10 | L/5 | L/2 | L | L/10 | L/5 | L/2 | L | L/10 | L/5 | L/2 | L |
| EVfold | 0.15 | 0.12 | 0.08 | 0.06 | 0.24 | 0.18 | 0.11 | 0.08 | 0.39 | 0.33 | 0.23 | 0.16 |
| PSICOV | 0.20 | 0.14 | 0.09 | 0.06 | 0.25 | 0.18 | 0.11 | 0.07 | 0.38 | 0.31 | 0.21 | 0.15 |
| CCMpred | 0.24 | 0.17 | 0.10 | 0.07 | 0.32 | 0.23 | 0.13 | 0.08 | 0.47 | 0.40 | 0.28 | 0.19 |
| MetaPSICOV | 0.40 | 0.31 | 0.19 | 0.12 | 0.43 | 0.34 | 0.23 | 0.15 | 0.55 | 0.49 | 0.37 | 0.27 |
| MP | 0.35 | 0.27 | 0.16 | 0.10 | 0.48 | 0.37 | 0.24 | 0.16 | 0.63 | 0.57 | 0.45 | 0.33 |
| NonMP | 0.51 | 0.39 | 0.24 | 0.14 | 0.58 | 0.46 | 0.29 | 0.18 | 0.70 | 0.66 | 0.55 | 0.41 |
| Mixed | 0.53 | 0.40 | 0.24 | 0.14 | 0.60 | 0.48 | 0.31 | 0.20 | 0.73 | 0.69 | 0.58 | 0.44 |
| Only contacts in transmembrane regions are evaluated | | | | | | | | | | | | |
| Method | Short | | | | Medium | | | | Long | | | |
| | L/10 | L/5 | L/2 | L | L/10 | L/5 | L/2 | L | L/10 | L/5 | L/2 | L |
| EVfold | 0.21 | 0.16 | 0.11 | 0.08 | 0.34 | 0.27 | 0.18 | 0.12 | 0.36 | 0.31 | 0.22 | 0.16 |
| PSICOV | 0.24 | 0.19 | 0.12 | 0.09 | 0.32 | 0.24 | 0.16 | 0.11 | 0.33 | 0.27 | 0.19 | 0.14 |
| CCMpred | 0.29 | 0.22 | 0.14 | 0.09 | 0.38 | 0.30 | 0.18 | 0.12 | 0.40 | 0.34 | 0.25 | 0.17 |
| MetaPSICOV | 0.41 | 0.32 | 0.22 | 0.16 | 0.46 | 0.38 | 0.26 | 0.18 | 0.45 | 0.39 | 0.29 | 0.22 |
| MP | 0.48 | 0.38 | 0.24 | 0.16 | 0.55 | 0.47 | 0.32 | 0.21 | 0.53 | 0.47 | 0.36 | 0.27 |
| NonMP | 0.54 | 0.42 | 0.27 | 0.18 | 0.58 | 0.50 | 0.34 | 0.23 | 0.57 | 0.53 | 0.42 | 0.31 |
| Mixed | 0.56 | 0.45 | 0.29 | 0.19 | 0.63 | 0.55 | 0.38 | 0.25 | 0.63 | 0.59 | 0.48 | 0.36 |

Fig. 2 (A-B) show the prediction accuracy with respect to the number of non-redundant sequence homologs (denoted as Meff) available for a protein under prediction. Our methods work particularly well

when a protein in question does not have a large Meff. Even if Meff>1000 (i.e., $\ln(Meff) \geq 9$), our method still outperforms MetaPSICOV and the pure co-evolution method CCMpred. This suggests that our deep learning method predicts contacts by exploiting extra information that is orthogonal to pairwise co-evolution strength generated by CCMpred.

## Contact-assisted folding accuracy

We build 3D structure models of an MP by feeding its top predicted contacts as distance restraints to CNS (Briinger et al., 1998), which folds proteins from a given set of distance and angle restraints. We also feed secondary structure predicted by RaptorX-Property(Wang et al., 2016a) as angle constraints to CNS. We generate 200 models for each test protein and then pick 5 models with the best NOE score. When the best of top 5 models are evaluated, as shown in Fig. 2(D), the average TMscore (RMSD in Å) of the 3D structure models built from our three model training strategies MP-only, NonMP-only and Mixed are 0.473 (14.4), 0.514 (12.7), and 0.548 (10.6), respectively. By contrast, the average TMscore (RMSD in Å) of the 3D structure models built from MetaPSICOV and CCMpred-predicted contacts are 0.41 (16.2) and 0.384 (16.5), respectively. Our three model training strategies can predict correct folds (i.e., TMscore≥0.6) for 113, 165, and 218 of 510 MPs, respectively, while MetaPSICOV and CCMpred can do so for only 77 and 56 of them, respectively. In terms of the number of models with TMscore>0.7 or 0.8, our methods have even a larger advantage over the others. In addition, our NonMP and Mixed strategies can produce 3D models with RMSD<6.5Å for 126 and 169 MPs, respectively, and 3D models with RMSD<4Å for 32 and 57 MPs, respectively. By contrast, CCMpred and MetaPSICOV can produce 3D models with RMSD<6.5Å for 49 and 64 MPs, respectively, and 3D models with RMSD<4Å for 10 and 14 MPs, respectively. Fig. 3 compares our three training strategies with CCMpred and MetaPSICOV in detail.

We have studied the 3D modeling accuracy with respect to the number of effective sequence homologs available for a protein in question. Our deep models outperform CCMpred and MetaPSICOV no matter how many sequence homologs are available (Figure 2C). Our NonMP and Mixed strategies can produce 3D models with TMscore at least 0.1 better than MetaPSICOV and CCMpred even when the protein in question has thousands of sequence homologs. This shows that the extra information used by our deep learning method is important not only for contact prediction, but also for 3D structure modeling.

We also compared our contact-assisted folding with template-based models (TBMs) built from the training proteins (Figure 2D). To compare with our NonMP strategy, we build TBMs for the 510 MPs from the 9627 non-MPs. To compare with our Mixed strategy, we build TBMs for the 510 MPs from a mix of the 9627 non-MPs and 4 of 5 MPs. For each test MP, we run HHpred (Söding, 2005) to search for the 5 best templates among the training proteins and then build 5 TBMs by MODELLER (Webb and Sali, 2014), which generates 3D structural models by mainly copying from templates. As shown in Fig. 2(D), our Mixed strategy on average has TMscore 0.548 and RMSD 10.6Å and produces 218 3D models with TMscore>0.6. By contrast, the corresponding TBMs have an average TMscore 0.35 and RMSD 17.2Å when only the first models are evaluated. When the best of the top 5 models are considered, the TBMs have an average TMscore 0.38 and RMSD 16.5Å. In total only 40 TBMs have TMscore>0.6, worse than our Mixed strategy. Our NonMP strategy on average has TMscore 0.514 and RMSD 12.7Å and produces 165 3D models with TMscore>0.6. By contrast, the corresponding TBMs have an average TMscore 0.14 and RMSD 126.3Å when the first models are considered, and the best of the top 5 TBMs have an average TMscore 0.17 and RMSD 111.5 Å. In total only 3 TBMs have TMscore>0.6 since there is almost no redundancy between our training non-MPs and the 510 MPs. Meanwhile, no TBMs have RMSD<4Å and only 2 TBMs have RMSD<6.5Å. All these results suggest that contact-assisted folding may work much better for MP modeling unless good templates are available in PDB.

## Blind test in CAMEO

We have implemented our algorithm as a fully-automated contact prediction web server (http://raptorx.uchicago.edu/ContactMap/) and blindly tested it through the weekly live benchmark CAMEO (http://www.cameo3d.org/) operated by the Schwede group(Haas et al., 2013a). By "blind" it means that the experimentally solved structure of a test protein has not been released in PDB when it is used as a test target. CAMEO(Haas et al., 2013b) can be interpreted as a fully-automated CASP(Moult et al., 2014), but has a smaller number (about 32) of participating servers since many CASP-participating servers are not fully automated. However, the CAMEO participants include some well-known servers such as Robetta(Kim et al., 2004) and its test version Server19(Ovchinnikov et al., 2017), Phyre(Kelley et al., 2015), RaptorX(Källberg et al., 2012), Swiss-Model(Biasini et al., 2014) and HHpred(Remmert et al., 2012). Meanwhile Robetta (Server19) employ both contact-assisted and fragment assembly-based ab initio folding and template-based modeling while the latter four mainly template-based modeling. Robetta predict contacts using a pure co-evolution method Gremlin(Kamisetty et al., 2013) enhanced by metagenomics data. With the same MSA, on average Gremlin has a similar contact prediction accuracy as CCMpred. Every weekend CAMEO releases 20 sequences for prediction test. The test proteins used by CAMEO have no publicly available native structures before it finishes collecting models from servers. Since experimentally solving the structures of MPs is challenging, starting from September 2016 to April 2017, we have observed only 4 MPs among all CAMEO hard targets: 5h36A, 5h35E, 5jkiA and 5l0wA. Since 5h56A and 5h35E are homologous, we describe the test results of 5h35E (CAMEO ID: 2017-01-07_00000030_3), 5jkiA(CAMEO ID: 2017-02-18_00000075_1) and 5l0wA (CAMEO ID: 2017-03-18_00000059_2). The CAMEO ID of our fully-automated server is Server60. For 5h36A and 5h35E predicted 3D models with RMSD close to 2Å and TMscore close to 0.9, better than all the other participating servers.

**Case study of 5h35E**

5h35E is an intracellular cation channel orthologue from *Sulfolobus solfataricus*, with 7 transmembrane helices, 212 residues and 1031 effective sequence homologs. Fig. 4 shows that even for a protein with so many sequence homologs, our server yields better contact prediction than CCMpred and MetaPSICOV. In particular, our predicted contact map has top L long-range accuracy 77.8%, while CCMpred- and MetaPSICOV-predicted contact maps have top L long-range accuracy 34.0% and 57.1%, respectively. Fig. 4 (B-D) show the overlap of top L predicted contacts with the native contact map, and that Evfold, CCMpred and MetaPSICOV have many more false positives.

The first 3D model submitted by our contact server has TMscore 0.86 and RMSD 2.59Å. The best of the five models submitted by our server has TMscore 0.88 and RMSD 2.29Å. All our 5 models have TMscore ≥0.83. Fig. 4 (E) shows the superimposition between our predicted model and the native structure. To examine the superimposition of our model with its native structure from various angles, please see http://raptorx.uchicago.edu/DeepAlign/44622802/. The best of top 5 models built by CNS from CCMpred and MetaPSICOV contacts have TMscore 0.756 and 0.771 and RMSD 3.76Å and 3.63Å, respectively. Among the top 5 models produced by CCMpred, only one of them has TMscore >0.75 while all the other 4 have TMscore<0.61. Similarly, all the other 4 models produced by MetaPSICOV have TMscore<0.65. When tested on the other MP target 5h36A, our method has similar quality as 5h35E, while CCMpred yields significantly worse 3D models. The best of top five 3D models built by CNS from CCMpred contacts for 5h36A has TMscore only 0.48.

Our server predicted better contact-assisted models than the other CAMEO-participating servers, including the top-notch servers Robetta and Server19, HHpred, RaptorX, SPARKS-X, and RBO Aleph (template-based and ab initio folding). Specifically, the best submitted model by the other servers has TMScore=0.69 and RMSD>5Å. The most structurally similar proteins in PDB70 found by DeepSearch (Eddy, 2011) for 5h35E are 5egiA and 5eikA, which have TMscore 0.677 and 0.653 with the native structure of this target, respectively. This is consistent with the fact that all the other CAMEO-participating servers predicted models with TMscore less than 0.7.

We have also submitted the sequence of 5h35E to the Evfold web server(Marks et al., 2011), which predicts contacts with the PLM option. Evfold detected more sequence homologs (i.e., Meff>2000). However, in terms of both contact prediction accuracy and 3D modeling accuracy, Evfold is worse than ours (Figure 4). In particular, the long-range top L/10, L/5, L/2 and L contact accuracy of Evfold is 0.81,

0.81, 0.642, and 0.524, respectively, better than CCMpred maybe because Evfold used many more sequence homologs. The first 3D model and the best of top 50 models produced by Evfold have TMscore 0.588 and 0.597 and RMSD 6.42Å and 6.31Å, respectively. This result further confirms that even for proteins with many sequence homologs, pure co-evolution methods still underperform our method because we predict contacts by exploiting extra information orthogonal to co-evolution information.

**Case study of 5jkiA**

5jkiA has 222 residues, 6 transmembrane helices and 1 amphiphilic helix, and 7244 effective sequence homologs. As shown in Suppl. Fig. 1, our server produced better contact prediction than CCMpred, MetaPSICOV and EVfold(web). Suppl. Fig. 1(B-D) show the overlap of top L predicted contacts with the native contact map. Evfold, CCMpred and MetaPSICOV have more false positives than ours. The best of the five 3D models submitted by our server has TMscore 0.77 and RMSD 3.55Å. The best of top 5 models built from CCMpred and MetaPSICOV contacts have TMscore 0.70 and 0.72, respectively, and RMSD 4.47 Å and 4.27Å. The best of top 5 models built by EVfold(web) has TMscore 0.61. To examine the superimposition of our model with its native structure from various angles, see http://raptorx.uchicago.edu/DeepAlign/15387452/. Our server predicted better 3D models than all the other CAMEO-participating servers, which submitted 3D models with TMscore<0.73 and RMSD>5.8 Å. This target shows that even when a protein under prediction has 7000 sequence homologs, our method still performs better.

**Case study of 5l0wA**

5l0wA is a post-translational translocation Sec71/Sec72 complex from *Escherichia coli* with 184 residues. Since it has only 100 effective sequence homologs, Evfold(web) did not return any prediction results. Our server produced better contact prediction than CCMpred and MetaPSICOV (Figure S2). Suppl. Fig. 2(B-C) shows the overlap of top L predicted contacts with the native contact map. Both CCMpred and MetaPSICOV have many more false positives. Our method also produces many more true positives than CCMpred. The first 3D model submitted by our server has TMscore 0.610 and RMSD 7.48Å. The best of top 5 models built from CCMpred and MetaPSICOV contacts have TMscore 0.417 and 0.555, respectively, and RMSD 20.71Å and 8.22Å. See http://raptorx.uchicago.edu/DeepAlign/99505283/ for the superimposition of our model with its native structure. Our server predicted better 3D models than all the other CAMEO-participating servers, which submitted 3D models with TMscore<0.43 and RMSD >30.0Å.

## Why does deep transfer learning work?

There is little sequence (profile) similarity between the 510 MPs and our training non-MPs, so the predictive power of our deep model trained by non-MPs is not from sequential features, but mostly from pairwise features. We conducted one experiment to measure the importance of sequential features, by training a new deep model of the same network architecture (in the 2D module) using only the non-MPs but without any sequential features. It turns out that this new deep model has almost the same accuracy as our deep model trained with sequential features included. The top L/k (k=1, 2, 5, 10) contact prediction accuracy decreases by about 1%. This result implies that our deep model does not rely much on sequence (profile) and secondary structure similarity to predict contacts. Instead, it predicts contacts by mainly learning contact patterns shared between non-MPs and MPs.

To further understand why our deep model learned from non-MPs works well on MP contact prediction, we examined the contact occurrence patterns in both MPs and non-MPs. To save space, here we focus only on all the 5×5 contact submatrices with at least 2 long-range contacts inside. In total there are $2^{25}$-26 (about 33 millions) of possible 5×5 contact patterns, but many do not appear in a protein contact map. The top 20 most frequent 5×5 contact patterns of both non-MPs and MPs are quite similar and differ mainly in the ranking order (Suppl. Fig. 3). The accumulative frequency of the top 20 5×5 contact patterns extracted from non-MPs and MPs are 11.6% and 12.8%, respectively, larger than the expected frequency (~$6.0×10^{-7}$). That is, the contact map of non-MPs and MPs is built from a set of basic building blocks with similar occurrence frequency, which justifies why we can improve contact prediction accuracy for MPs by learning from non-MPs.

To further study this problem, we have trained several deep models by 510 randomly chosen non-MPs. Such models have top L/10, L/5, L/2 and L long-range accuracy about 0.57, 0.51, 0.39 and 0.29, respectively, and top L/10, L/5, L/2 and L medium-range accuracy 0.42, 0.34, 0.23 and 0.15, respectively. In contrast, the model trained by only MPs has top L/10, L/5, L/2 and L long-range accuracy 0.63, 0.57, 0.45 and 0.33, respectively, and top L/10, L/5, L/2 and L medium-range accuracy 0.35, 0.27, 0.16 and 0.10, respectively. That is, the model trained by 510 non-MPs has worse performance than that trained by a similar number of MPs, but the gap is not large. This further confirms that MPs and non-MPs indeed share common properties in contacts, but the set of 510 non-MPs is not enough to cover all the possible patterns in existing MPs.

## Study of human membrane proteins

Human transmembrane proteins play an important role in the living cells for energy production, regulation and metabolism (Kozma et al., 2012). The fact that half of drugs have some effect on human transmembrane proteins also underlines their biological importance (Kozma et al., 2012). In total there are 5182 reviewed human MPs in UnitProt. It was estimated that at least ~12% of the Human genome might encode multi-pass transmembrane proteins according to UniProt(2015) (i.e., ~2500 multi-pass transmembrane proteins). Due to the structural and physiochemical properties of these proteins, experimentally determining their structures is challenging.

Here we study for how many human MPs our deep learning method can predict a correct fold (TMscore≥0.5 or 0.6). Since we are mostly interested in multi-pass MPs, we exclude single-pass MPs from consideration. After removing one-pass human MPs and those MPs with more than 700 amino acids (CCMpred crashes on such a large protein due to GPU memory limit), we have 2215 reviewed human multi-pass MPs for study. In order to obtain a good estimation, we first establish the relationship between 3D model quality and the number of effective sequence homologs (i.e., Meff) available for a protein under prediction and its sequence length, using the 354 of the 510 multi-pass MPs with solved structures. Fig. 6 shows the TMscore of the 354 MPs with respect to the length-normalized Meff, denoted as Neff. We calculate Neff by $\frac{Meff}{(Len)^{0.7}}$. Fig. 5(A) shows that $ln(Neff)$ has good correlation with TMscore. When $ln(Neff)$ is larger than 1.5 and 3.5, the predicted models on average have TMscore≥0.5 and TMscore≥0.6, respectively. That is, for a protein with 150, 200, 250, 300, 350, 400, 450, and 500 residues, we will need 150, 183, 214, 242, 270, 297, 322, and 347 effective sequence homologs, respectively, to predict a model with TMscore>0.5 and 1105, 1351, 1580, 1795, 1999, 2195, 2384 and 2566 sequence homologs, respectively, to predict a model with TMscore>0.6. Based upon this, we may estimate that our deep method can fold 1871 and 1345 human multi-pass MPs with TMscore ≥ 0.5 and 0.6, respectively. See Fig. 5(B) and (C) for the Neff distribution of the 354 multi-pass MPs and the 2215 reviewed human multi-pass MPs. If we remove redundancy at 25% sequence identity level, there are about 1245 non-redundant MPs left, of which 951 and 555 can be predicted with TMscore ≥0.5 and 0.6, respectively. Excluding those in the 1245 non-redundant MPs which have sequence identity >25% or BALST E-value<0.1 with any proteins in PDB, we have 666 MPs left, among which our method can predict 3D models with TMscore ≥0.5 and 0.6 for 420 and 188 MPs, respectively. That is, among these 2215 reviewed human MPs, our method may be able to successfully predict 188~420 new folds.

## Discussion

As opposed to existing supervised methods that predict contacts individually, our method predicts all contacts of a protein simultaneously, which makes it easy to exploit contact occurrence patterns. By learning from non-MPs, we can significantly improve MP contact prediction and accordingly folding. Even without any MPs in the training set, our deep learning model still performs well, outperforming the models trained by a limited number of MPs and existing best methods. Even for proteins with thousands of sequence homologs, our method still outperforms pure co-evolution methods. This is because our method predicts contacts by mainly exploiting contact occurrence patterns instead of sequence-level similarity, and that this kind of higher-order correlation is shared by non-MPs and MPs and orthogonal to pairwise co-evolution information.

Our method works better than pure co-evolution methods not by simply filling in more contacts between secondary structure elements, as evidenced by the fact that even if only top L/10 (which is not a big number) are evaluated, our method still outperforms pure co-evolution methods. Our method works mostly because it has fewer false positives especially when the protein under prediction does not have many sequence homologs. See section "Blind test in CAMEO" for the contact map overlap of three concrete examples. Our predicted contacts are also widely distributed across different secondary structure elements.

Large-scale study on the 510 non-redundant MPs with solved structures in PDB and the human genome shows that our method can produce correct folds for 40-60% of the 510 MPs and 60-80% of the reviewed human multi-pass MPs, which to our knowledge has never been reported before. This is significant since it is challenging to solve MP structures by experimental techniques and homology modeling. Further, our method is of low cost and high throughput. For one test MP, sequence homologs can be detected in minutes, contact prediction can be done in seconds on a single GPU card (~$1200) and 3D structure models can be built within from 30 minutes to a few hours on a Linux workstation of 20 CPUs (~$6000). Currently we build a 3D model of a test protein using a simple way. It is possible to further improve 3D modeling accuracy by combining our predicted contacts with fragment assembly and some MP-specific topology information.

By adding MP-specific features such as lipid accessibility (Phatak et al., 2011) and topology structure (Tsirigos et al., 2015) to our deep learning models, we can improve contact accuracy by about 1%. This might be because our deep learning models have already implicitly learned them. To simplify the presentation, the results presented here are derived from the models without MP-specific features. We have also tried other deep transfer learning strategies. For example, we extract the output of the last convolutional layer of two deep models MP-only and NonMP-only, concatenate them into a single feature set and then use this new feature set to train a 2-layer fully connected neural network for MP contact prediction. However, this transfer learning strategy cannot outperform our NonMP strategy.

One of the major drawbacks with deep learning is that it is hard to interpret the resultant model, although the machine learning community has been actively working on this. It is not easy to extract simple rules from our deep model to explain why a specific residue pair is predicted to form a contact. Otherwise we may develop a simpler model for contact prediction. However, we have studied what has been learned by the hidden neurons in the last layer of the 2D residual network, which have 75 hidden neurons. We fed all the non-MPs to this deep network. At each residue pair, we focus on those neurons with the largest activation values (larger than its mean plus twice the standard deviation). These neurons dominate the others in predicting if this residue pair forms a contact or not. We can interpret that each of these neurons represents the small contact submatrix centered at this residual pair. For simplicity, here we consider only the 3×3 submatrix. By clustering all the 3×3 contact submatrices represented by one neuron, we can find out the contact patterns learned by this neuron. It turns out that the set of contact submatrices represented by a single neuron are quite similar. Their average hamming distance is substantially smaller than the expected value of any two randomly chosen submatrices. This implies that each hidden neuron in the last layer indeed represents a small number of contact occurrence patterns. Similarly, we may obtain the contact patterns derived from the 510 MPs for each of the 75 hidden neurons. Again, we find out that the set of submatrices for each hidden neuron are quite similar and have a small hamming distance. Further, the contact patterns learned by a specific hidden neuron from non-MPs are quite similar to those learned from MPs. This explains why the deep model trained by non-MPs performs well in predicting contacts of MPs.

## Author contributions

J.X. conceived the project, developed the algorithm and wrote the paper. S.W. did data preparation and analysis, built three-dimensional models from predicted contacts and helped with paper writing (in Chicago) and revised the paper while being affiliated with KAUST. Z.L. helped with algorithm development and data analysis. Y.Y. helped with algorithm development and proofread the paper.

## Acknowledgements

This work is supported by National Institutes of Health grant R01GM089753 to J.X. and National Science Foundation grant DBI-1564955 to J.X. The authors are also grateful to the support of Nvidia Inc. and the computational resources provided by XSEDE.

# References

ADHIKARI, B., BHATTACHARYA, D., CAO, R. & CHENG, J. (2015). CONFOLD: residue - residue contact - guided ab initio protein folding. *Proteins: Structure, Function, and Bioinformatics,* 83, 1436-1449.

ALTSCHUL, S. F., MADDEN, T. L., SCHÄFFER, A. A., ZHANG, J., ZHANG, Z., MILLER, W. & LIPMAN, D. J. (1997). Gapped BLAST and PSI-BLAST: a new generation of protein database search programs. *Nucleic acids research,* 25, 3389-3402.

BETANCOURT, M. R. & THIRUMALAI, D. (1999). Pair potentials for protein folding: choice of reference states and sensitivity of predicted native states to variations in the interaction schemes. *Protein Science,* 8, 361-369.

BIASINI, M., BIENERT, S., WATERHOUSE, A., ARNOLD, K., STUDER, G., SCHMIDT, T., KIEFER, F., CASSARINO, T. G., BERTONI, M., BORDOLI, L. & SCHWEDE, T. (2014). SWISS-MODEL: modelling protein tertiary and quaternary structure using evolutionary information. *Nucleic Acids Research,* 42, W252-W258.

BRIINGER, A. T., ADAMS, P. D., CLORE, G. M., DELANO, W. L., GROS, P., GROSSE-KUNSTLEVE, R. W., JIANG, J.-S., KUSZEWSKI, J., NILGES, M. & PANNU, N. S. (1998). Crystallography & NMR system: A new software suite for macromolecular structure determination. *Acta Crystallogr D Biol Crystallogr,* 54, 905-921.

BRUNGER, A. T. (2007). Version 1.2 of the Crystallography and NMR system. *Nature protocols,* 2, 2728-2733.

CONSORTIUM, U. (2015). UniProt: a hub for protein information. *Nucleic Acids Res,* 43, D204-12.

DI LENA, P., NAGATA, K. & BALDI, P. (2012). Deep architectures for protein contact map prediction. *Bioinformatics,* 28, 2449-2457.

EDDY, S. R. (2011). Accelerated Profile HMM Searches. *PLoS Comput Biol,* 7, e1002195.

FUCHS, A., KIRSCHNER, A. & FRISHMAN, D. (2009). Prediction of helix–helix contacts and interacting helices in polytopic membrane proteins using neural networks. *Proteins: Structure, Function, and Bioinformatics,* 74, 857-871.

HAAS, J., ROTH, S., ARNOLD, K., KIEFER, F., SCHMIDT, T., BORDOLI, L. & SCHWEDE, T. (2013a). The Protein Model Portal--a comprehensive resource for protein structure and model information. *Database (Oxford),* 2013, bat031.

HAAS, J., ROTH, S., ARNOLD, K., KIEFER, F., SCHMIDT, T., BORDOLI, L. & SCHWEDE, T. (2013b). The Protein Model Portal—a comprehensive resource for protein structure and model information. *Database,* 2013, bat031.

HE, K., ZHANG, X., REN, S. & SUN, J. (2015). Deep residual learning for image recognition. *arXiv preprint arXiv:1512.03385*.

IOFFE, S. & SZEGEDY, C. (2015). Batch normalization: Accelerating deep network training by reducing internal covariate shift. *arXiv preprint arXiv:1502.03167*.

JONES, D. T., BUCHAN, D. W., COZZETTO, D. & PONTIL, M. (2012). PSICOV: precise structural contact prediction using sparse inverse covariance estimation on large multiple sequence alignments. *Bioinformatics,* 28, 184-190.

JONES, D. T., SINGH, T., KOSCIOLEK, T. & TETCHNER, S. (2015). MetaPSICOV: combining coevolution methods for accurate prediction of contacts and long range hydrogen bonding in proteins. *Bioinformatics,* 31, 999-1006.

KÄLLBERG, M., WANG, H., WANG, S., PENG, J., WANG, Z., LU, H. & XU, J. (2012). Template-based protein structure modeling using the RaptorX web server. *Nature protocols,* 7, 1511-1522.

KAMISETTY, H., OVCHINNIKOV, S. & BAKER, D. (2013). Assessing the utility of coevolution-based residue–residue contact predictions in a sequence-and structure-rich era. *Proceedings of the National Academy of Sciences,* 110, 15674-15679.

KELLEY, L. A., MEZULIS, S., YATES, C. M., WASS, M. N. & STERNBERG, M. J. E. (2015). The Phyre2 web portal for protein modeling, prediction and analysis. *Nature protocols,* 10, 845-858.

KIM, D. E., CHIVIAN, D. & BAKER, D. (2004). Protein structure prediction and analysis using the Robetta server. *Nucleic Acids Research,* 32, W526-W531.

KOZMA, D., SIMON, I. & TUSNADY, G. E. (2012). PDBTM: Protein Data Bank of transmembrane proteins after 8 years. *Nucleic acids research,* 41, D524-9.

KROGH, A., LARSSON, B., VON HEIJNE, G. & SONNHAMMER, E. L. (2001). Predicting transmembrane protein topology with a hidden Markov model: application to complete genomes. *Journal of molecular biology,* 305, 567-580.

LACAPERE, J.-J., PEBAY-PEYROULA, E., NEUMANN, J.-M. & ETCHEBEST, C. (2007). Determining membrane protein structures: still a challenge! *Trends in biochemical sciences,* 32, 259-270.

LIN, T.-Y., MAIRE, M., BELONGIE, S., HAYS, J., PERONA, P., RAMANAN, D., DOLLÁR, P. & ZITNICK, C. L. (Year). Microsoft coco: Common objects in context. *In:* European Conference on Computer Vision, 2014. Springer, 740-755.

LO, A., CHIU, Y.-Y., RØDLAND, E. A., LYU, P.-C., SUNG, T.-Y. & HSU, W.-L. (2009). Predicting helix–helix interactions from residue contacts in membrane proteins. *Bioinformatics,* 25, 996-1003.

MA, J., WANG, S., WANG, Z. & XU, J. (2015). Protein contact prediction by integrating joint evolutionary coupling analysis and supervised learning. *Bioinformatics,* 31, 3506-13.

MARKS, D. S., COLWELL, L. J., SHERIDAN, R., HOPF, T. A., PAGNANI, A., ZECCHINA, R. & SANDER, C. (2011). Protein 3D structure computed from evolutionary sequence variation. *PloS one,* 6, e28766.

MCALLISTER, S. R. & FLOUDAS, C. A. (2008). α-Helical topology prediction and generation of distance restraints in membrane proteins. *Biophysical journal,* 95, 5281-5295.

MIYAZAWA, S. & JERNIGAN, R. L. (1985). Estimation of effective interresidue contact energies from protein crystal structures: quasi-chemical approximation. *Macromolecules,* 18, 534-552.

MONASTYRSKYY, B., D'ANDREA, D., FIDELIS, K., TRAMONTANO, A. & KRYSHTAFOVYCH, A. (2015). New encouraging developments in contact prediction: Assessment of the CASP11 results. *Proteins: Structure, Function, and Bioinformatics*.

MOULT, J., FIDELIS, K., KRYSHTAFOVYCH, A., SCHWEDE, T. & TRAMONTANO, A. (2014). Critical assessment of methods of protein structure prediction (CASP)--round x. *Proteins,* 82 Suppl 2, 1-6.

NAIR, V. & HINTON, G. E. (Year). Rectified linear units improve restricted boltzmann machines. *In:* Proceedings of the 27th International Conference on Machine Learning (ICML-10), 2010. 807-814.

NUGENT, T. & JONES, D. T. (2010). Predicting transmembrane helix packing arrangements using residue contacts and a force-directed algorithm. *PLoS Comput Biol,* 6, e1000714.

OVCHINNIKOV, S., PARK, H., VARGHESE, N., HUANG, P.-S., PAVLOPOULOS, G. A., KIM, D. E., KAMISETTY, H., KYRPIDES, N. C. & BAKER, D. (2017). Protein structure determination using metagenome sequence data. *Science,* 355, 294-298.

PHATAK, M., ADAMCZAK, R., CAO, B., WAGNER, M. & MELLER, J. (2011). Solvent and lipid accessibility prediction as a basis for model quality assessment in soluble and membrane proteins. *Current Protein and Peptide Science,* 12, 563-573.

REMMERT, M., BIEGERT, A., HAUSER, A. & SÖDING, J. (2012). HHblits: lightning-fast iterative protein sequence searching by HMM-HMM alignment. *Nature methods,* 9, 173-175.

RUSSAKOVSKY, O., DENG, J., SU, H., KRAUSE, J., SATHEESH, S., MA, S., HUANG, Z., KARPATHY, A., KHOSLA, A. & BERNSTEIN, M. (2015). Imagenet large scale visual recognition challenge. *International Journal of Computer Vision,* 115, 211-252.

SEEMAYER, S., GRUBER, M. & SÖDING, J. (2014). CCMpred—fast and precise prediction of protein residue–residue contacts from correlated mutations. *Bioinformatics,* 30, 3128-3130.

SKWARK, M. J., MICHEL, M., HURTADO, D. M., EKEBERG, M. & ELOFSSON, A. (2016). Predicting accurate contacts in thousands of Pfam domain families using PconsC3. *Bioinformatics*.

SKWARK, M. J., RAIMONDI, D., MICHEL, M. & ELOFSSON, A. (2014). Improved contact predictions using the recognition of protein like contact patterns. *PLoS computational biology,* 10, e1003889.

SÖDING, J. (2005). Protein homology detection by HMM–HMM comparison. *Bioinformatics,* 21, 951-960.

TSIRIGOS, K. D., PETERS, C., SHU, N., KÄLL, L. & ELOFSSON, A. (2015). The TOPCONS web server for consensus prediction of membrane protein topology and signal peptides. *Nucleic acids research*, gkv485.

UHLÉN, M., FAGERBERG, L., HALLSTRÖM, B. M., LINDSKOG, C., OKSVOLD, P., MARDINOGLU, A., SIVERTSSON, Å., KAMPF, C., SJÖSTEDT, E. & ASPLUND, A. (2015). Tissue-based map of the human proteome. *Science,* 347, 1260419.

WALLIN, E. & HEIJNE, G. V. (1998). Genome - wide analysis of integral membrane proteins from eubacterial, archaean, and eukaryotic organisms. *Protein Science,* 7, 1029-1038.

WANG, G. & DUNBRACK, R. L. (2003). PISCES: a protein sequence culling server. *Bioinformatics,* 19, 1589-1591.

WANG, S., LI, W., LIU, S. & XU, J. (2016a). RaptorX-Property: a web server for protein structure property prediction. *Nucleic acids research,* 44, W430-435.

WANG, S., PENG, J., MA, J. & XU, J. (2016b). Protein secondary structure prediction using deep convolutional neural fields. *Scientific reports,* 6.

WANG, S., SUN, S., LI, Z., ZHANG, R. & XU, J. (2017). Accurate De Novo Prediction of Protein Contact Map by Ultra-Deep Learning Model. *PLoS Comput Biol,* 13, e1005324.

WANG, X.-F., CHEN, Z., WANG, C., YAN, R.-X., ZHANG, Z. & SONG, J. (2011). Predicting residue-residue contacts and helix-helix interactions in transmembrane proteins using an integrative feature-based random forest approach. *PloS one,* 6, e26767.

WANG, Z. & XU, J. (2013). Predicting protein contact map using evolutionary and physical constraints by integer programming. *Bioinformatics,* 29, i266-i273.

WEBB, B. & SALI, A. (2014). Comparative protein structure modeling using Modeller. *Current protocols in bioinformatics*, 5.6. 1-5.6. 32.

WEIGT, M., WHITE, R. A., SZURMANT, H., HOCH, J. A. & HWA, T. (2009). Identification of direct residue contacts in protein-protein interaction by message passing. *Proc Natl Acad Sci U S A,* 106, 67-72.

WU, S. & ZHANG, Y. (2008). A comprehensive assessment of sequence-based and template-based methods for protein contact prediction. *Bioinformatics,* 24, 924-931.

YANG, J., JANG, R., ZHANG, Y. & SHEN, H.-B. (2013). High-accuracy prediction of transmembrane inter-helix contacts and application to GPCR 3D structure modeling. *Bioinformatics,* 29, 2579-87.

YıLDıRıM, M. A., GOH, K.-I., CUSICK, M. E., BARABÁSI, A.-L. & VIDAL, M. (2007). Drug—target network. *Nature biotechnology,* 25, 1119-1126.

ZHANG, H., HUANG, Q., BEI, Z., WEI, Y. & FLOUDAS, C. A. (2016a). COMSAT: Residue contact prediction of transmembrane proteins based on support vector machines and mixed integer linear programming. *Proteins: Structure, Function, and Bioinformatics,* 84, 332-348.

ZHANG, L., WANG, H., YAN, L., SU, L. & XU, D. (2016b). OMPcontact: An Outer Membrane Protein Inter-Barrel Residue Contact Prediction Method. *Journal of Computational Biology,* 24, 217-228.

ZHANG, Y. & SKOLNICK, J. (2004). Scoring function for automated assessment of protein structure template quality. *Proteins: Structure, Function, and Bioinformatics,* 57, 702-710.

# Figure Legend

**Figure 1.** Overview of our deep learning model for MP contact prediction where L is the sequence length of one MP under prediction.

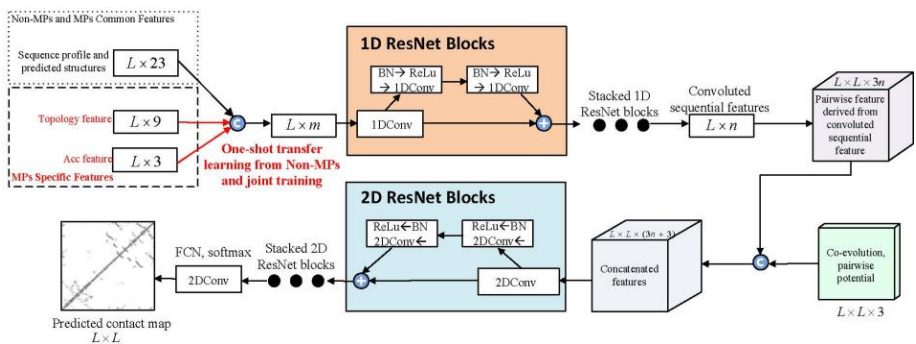

**Figure 2.** Top L/5 long-range accuracy **(A)**, medium-range accuracy **(B)**, and TMscore of the best of top 5 3D models **(C)** generated by our three models Mixed (cyan), NonMP-only (purple), MP-only (green) and CCMpred (blue) and MetaPSICOV (red) with respect to *ln(Meff)*. **(D)** Summary results of all tested methods in terms of modeling accuracy. Column '#<XÅ' lists the number of MPs whose 3D models have RMSD≤X Å. Column '#TM>Y' lists the number of MPs whose 3D models have TMscore≥ Y. $\overline{RMSD}$ ($\overline{TMsco}$) shows the average TMSD (TMscore) of all 510 MPs. TBM(MP) and TBM(NonMP) stands for template-based modeling with membrane proteins as templates and without membrane proteins as templates, respectively.

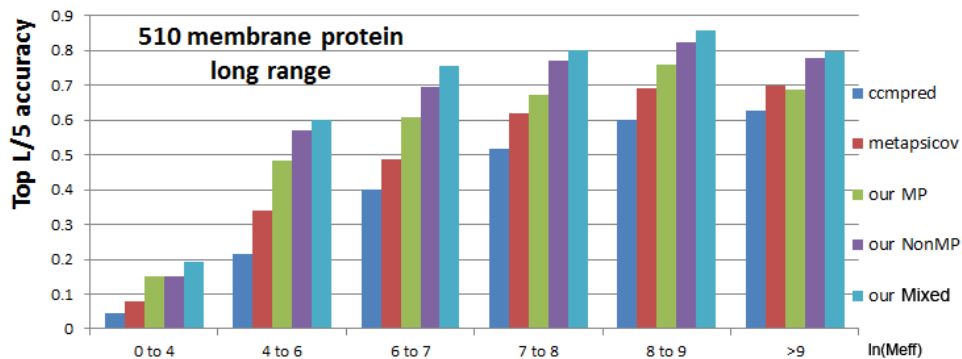

(A)

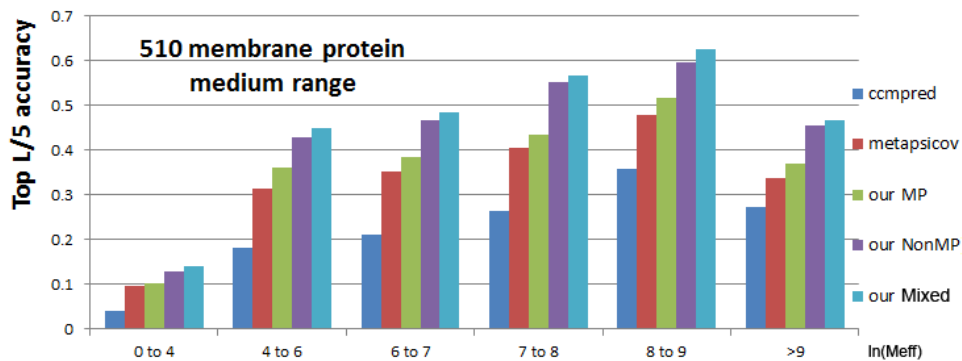

(B)

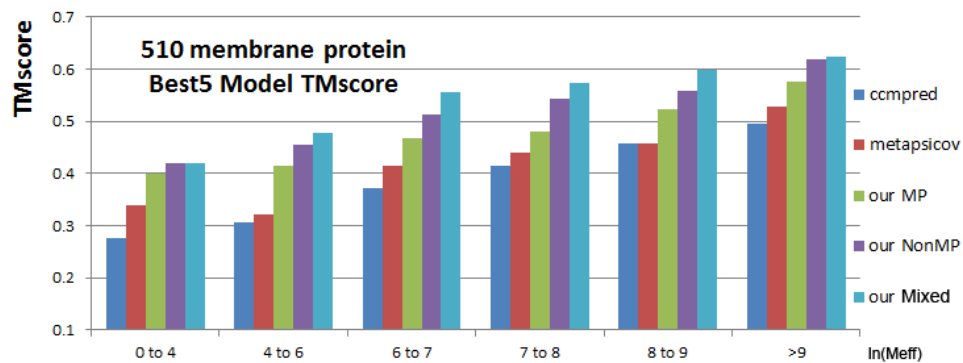

(C)

| Method | #<3Å | #<4Å | #<5Å | #<6.5Å | RMSD | TMsco | #TM>0.5 | #TM>0.6 | #TM>0.7 | #TM>0.8 |
|---|---|---|---|---|---|---|---|---|---|---|
| CCMpred | 1 | 10 | 18 | 49 | 16.5 | 0.384 | 122 | 56 | 15 | 1 |
| MetaPSICOV | 2 | 14 | 31 | 64 | 16.2 | 0.413 | 147 | 77 | 25 | 0 |
| TBM (NonMP) | 0 | 0 | 0 | 2 | 111.5 | 0.170 | 8 | 3 | 0 | 0 |
| TBM (MP) | 5 | 18 | 37 | 89 | 16.5 | 0.378 | 134 | 40 | 19 | 1 |
| MP | 4 | 16 | 41 | 101 | 14.4 | 0.473 | 212 | 113 | 48 | 3 |
| NonMP | 10 | 38 | 80 | 138 | 11.9 | 0.521 | 262 | 168 | 75 | 17 |
| Mixed | 20 | 57 | 108 | 169 | 10.6 | 0.548 | 288 | 218 | 120 | 33 |

(D)

**Figure 3.** Quality comparison of the best of top 5 contact-assisted models generated by our two methods, CCMpred and MetaPSICOV. **(A)** Mixed vs. CCMpred; **(B)** Mixed vs. MetaPSICOV; **(C)** NonMP vs. CCMpred; **(D)** NonMP vs. MetaPSICOV.

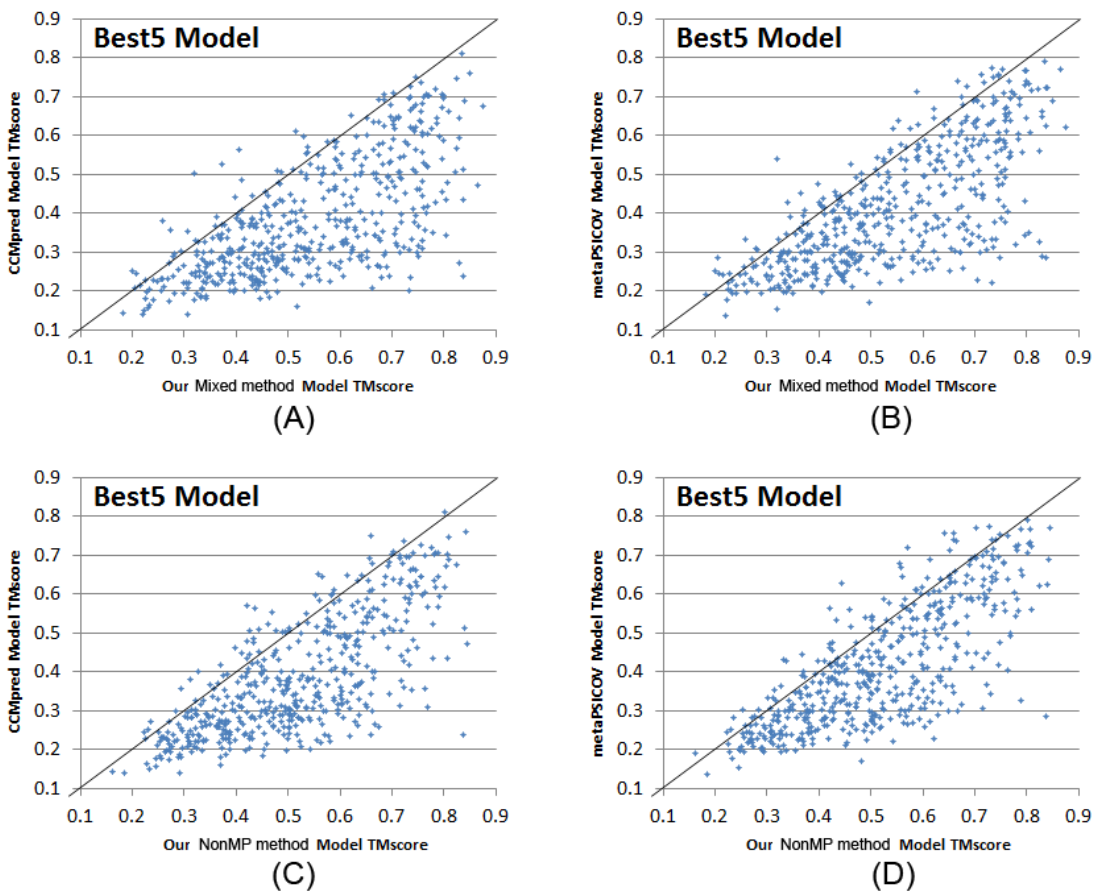

**Figure 4.** Case study of one CAMEO target 5h35E. (A) The long- and medium-range contact prediction accuracy of our methods, MetaPSICOV, CCMpred, and EVfold (web server). (B-D) The overlap between the native contact map and contact maps predicted by our method, CCMpred, MetaPSICOV, and EVfold. Top L predicted all-range contacts are displayed. A grey, red and green dot represents a native contact, a correct prediction and a wrong prediction, respectively. (E) The superimposition between our predicted model (in red) and the native structure (in green).

|  | Long range accuracy | | | | Medium range accuracy | | | |
|---|---|---|---|---|---|---|---|---|
|  | L | L/2 | L/5 | L/10 | L | L/2 | L/5 | L/10 |
| Our method | 0.778 | 0.953 | 1.000 | 1.000 | 0.316 | 0.547 | 0.905 | 1.000 |
| metaPSICOV | 0.571 | 0.774 | 0.929 | 1.000 | 0.245 | 0.401 | 0.619 | 0.810 |
| CCMpred | 0.340 | 0.528 | 0.786 | 0.810 | 0.137 | 0.217 | 0.429 | 0.619 |
| EVfold (web) | 0.425 | 0.642 | 0.810 | 0.810 | 0.160 | 0.283 | 0.524 | 0.762 |

(A)

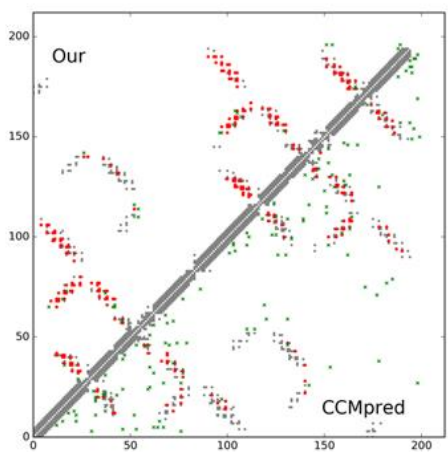

(B)

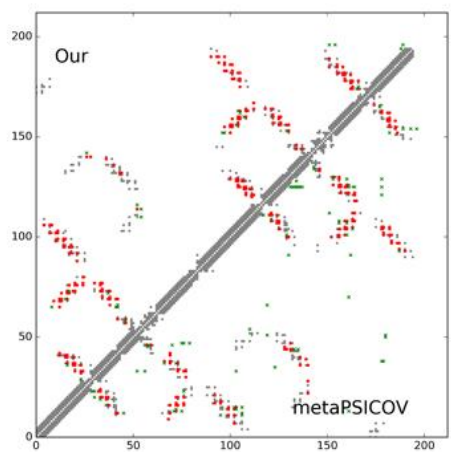

(C)

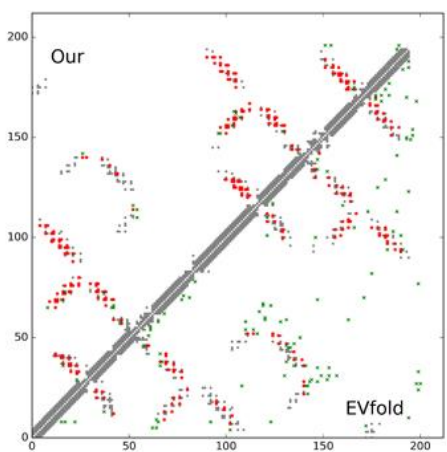

(D)

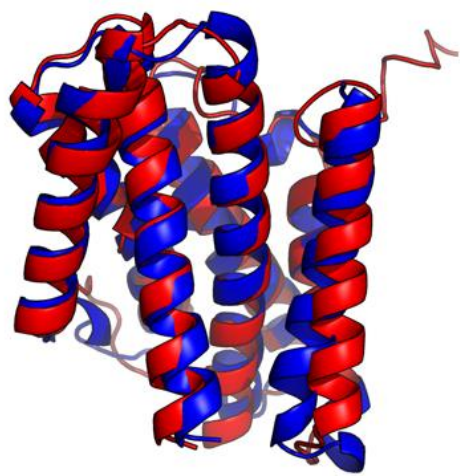

(E)

**Figure 5. (A)** TMscore with respect to *ln(Neff)*, based upon the 354 multi-pass membrane proteins in PDB. **(B)** *ln(Neff)* distribution of the 354 multi-pass MPs in PDB. **(C)** *ln(Neff)* distribution of the 2215 reviewed human multi-pass MPs.

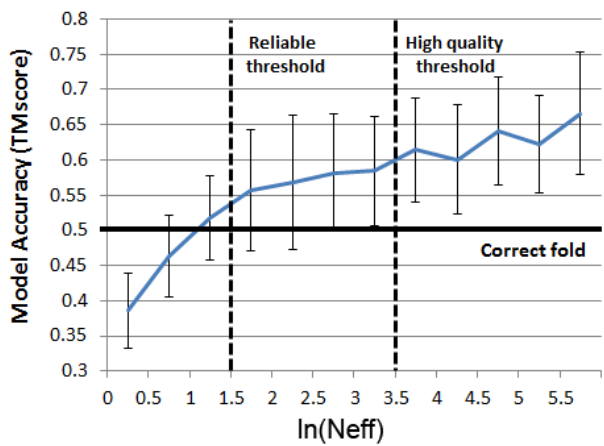

(A)

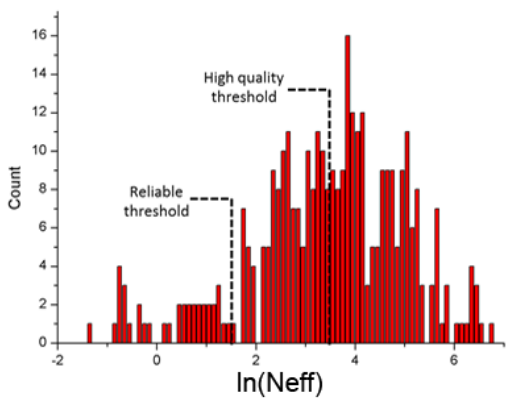

(B)

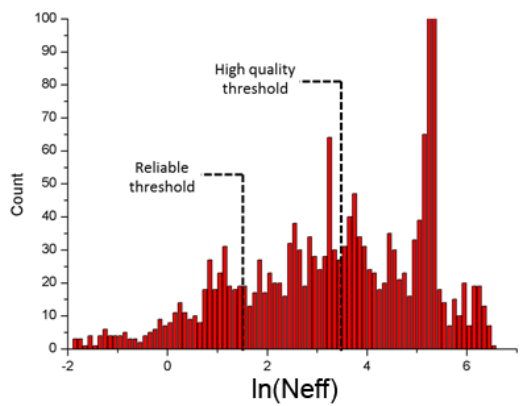

(C)

## STAR Methods

**METHOD DETAILS**

**Protein features**

Given a membrane protein (MP) sequence under prediction, we run PSI-BLAST (Altschul et al., 1997) or HHblits (Remmert et al., 2012) to find its sequence homologs and then build a multiple sequence alignment (MSA) of them. Starting from the MSA, we derive two types of protein features: sequential features and pairwise features, which are also called 1-dimensional (1D) and 2-dimensional (2D) features, respectively. The sequential features include protein sequence profile and secondary structure predicted by RaptorX-Property (Wang et al., 2016b, Wang et al., 2016a). The pairwise features include co-evolution strength generated by CCMpred (Seemayer et al., 2014), mutual information and pairwise contact potential (Miyazawa and Jernigan, 1985, Betancourt and Thirumalai, 1999). We have also tested some MP-specific features such as lipid accessibility and topology information.

**Calculation of Meff and Neff**

Meff measures the amount of homologous information in an MSA (multiple sequence alignment). It can be interpreted as the number of non-redundant (or effective) sequence homologs in an MSA when 70% sequence identity is used as cutoff. To calculate Meff, we first calculate the sequence identity between any two proteins in the MSA. Let a binary variable $S_{ij}$ denote the similarity between two protein sequences i and j. $S_{ij}$ is equal to 1 if and only if the sequence identity between i and j is at least 70%. For a protein i, we calculate the sum of $S_{ij}$ over all the proteins (including itself) in the MSA and denote it as $S_i$. Finally, we calculate Meff as the sum of $1/S_i$ over all the protein sequences in this MSA. Neff is length-normalized Meff. We calculate Neff by $\frac{Meff}{(Len)^{0.7}}$ to maximize the correlation between TMscore and Neff.

**Deep learning models and parameter optimization**

For the implementation of deep residual network, please see https://github.com/KaimingHe/deep-residual-networks. We train our deep learning models by maximum-likelihood with $L_2$-norm regularization, and use a stochastic gradient descent algorithm to minimize the objective function. We implement the whole algorithm with Theano (http://www.deeplearning.net/software/theano/) and run it on GPU.

**Data for model parameter optimization and test**

In total there are 510 non-redundant MPs with solved structures in PDBTM (Kozma et al., 2012), a database depositing all MPs with solved structures. We use them to evaluate prediction methods. Among these 510 MPs, 5 of them have no contacts in transmembrane regions and thus, are not used to evaluate contact prediction accuracy in transmembrane regions. When MPs are involved in training, we randomly divide the 510 MPs into 5 subsets of same size for 5-fold cross validation. That is, we use 4 subsets of MPs for training and the remaining 1 subset of MPs for test.

So far there are more than 10,000 proteins in PDB25, a representative set of non-redundant proteins in PDB in which any two proteins share less than 25% sequence identity (Wang and Dunbrack, 2003). To remove redundancy between MPs and PDB25, we exclude the proteins in PDB25 sharing >25% sequence identity or having a BLAST E-value <0.1 with any of the 510 MPs. This results in 9627 non-MPs in PDB25, all of which are non-redundant to the 510 MPs. From these 9627 non-MPs, we randomly sample 600 proteins as the validation set and use the remaining ~9000 proteins as the training proteins.

**Three model training strategies**

1) Training deep models by MPs only (denoted as MP-only). That is, we train our deep learning model using only the 510 MPs. In total we have trained 5 models and for each of them, we use 4/5 of the 510 (i.e., 408) MPs as the training and validation proteins and the remaining 1/5 MPs as the test proteins. We construct a validation set by randomly selecting 30 out of the 408 MPs. To reduce bias introduced by these 30 MPs, we also include 100 non-MPs in the validation set. The validation set is used to determine when to stop training and the regularization factor. Since there are only 378 MPs in the training set, we cannot use a very deep learning model. We tested several network architectures and found out that a deep model with two 1D convolutional layers (and 50 hidden neurons at each layer) and twenty 2D convolutional layers (and 30 hidden neurons at each layer) yields the best performance. We use the same architecture for all the 5 deep models and terminate the training procedure at 15 epochs (each epoch scans through all the training data once).

2) Training deep models by non-MPs only (denoted as NonMP-only). That is, we train our deep learning model without using any MPs. In this case, we have trained only one model using the ~9000 non-MPs and validate it using 600 non-MPs and tested it using all the 510 MPs. Since we have a large training set, we use a model with 6 1D convolutional layers (50 hidden neurons at each layer) and 60 2D convolutional layers (and 60 hidden neurons at each layer). We terminate the training procedure at 20 epochs.

3) Training deep models by a mix of non-MPs and MPs (denoted as Mixed). In total we have trained 5

models. Each model is trained by a mix of 9000 non-MPs and 4/5 of the MPs. We validate this model using the 600 non-MPs and test it by the remaining 1/5 of the MPs. Here we use the same network architecture as the NonMP-only strategy. Since there are fewer MPs than non-MPs, we assign a weight factor to each training MP to achieve the maximal accuracy. We have tried a few different weight factors for MPs, such as 1, 3, 5, 7, and 9. Our experimental results show that by setting the weight of MPs to 5 or 7, we can obtain better accuracy than the other values, so finally we set the weight factor to 5. By the way, when the weight of MPs is infinity, this strategy becomes the MP-only strategy. Again we terminate the training procedure at 20 epochs.

**Contact-assisted folding**

For each test protein, we feed top predicted contacts and predicted secondary structure to CNS, a software package that builds 3D models from distance and angle restraints, to predict the 3D models.

**Template-based modeling (TBM)**

To generate template-based models (TBMs) for a test protein, we first run HHblits (with the UniProt20_2016 library) to generate an HMM file for the test protein, then run HHsearch with this HMM file to search for the best templates among the training proteins of our deep learning model, and finally run MODELLER to build a TBM from each of the top 5 templates.

**Competing methods**

We have tested Evfold(web server) (Marks et al., 2011), CCMpred (Seemayer et al., 2014), and MetaPSICOV (Jones et al., 2015). The first two methods are representative co-evolution analysis methods. MetaPSICOV is a supervised learning method that performed the best in CASP11 (Monastyrskyy et al., 2015). These programs are run with parameters set according to their respective papers. We did not evaluate PconsC2 (Skwark et al., 2014) and its improved version PConsC3 since they are slow and did not outperform MetaPSICOV in latest CASPs (Monastyrskyy et al., 2015). It is challenging to evaluate MP-specific tools because of the following reasons. First, some tools such as OMPcontact (Zhang et al., 2016b) are not available. Second, some tools need extra input information, such as [24]. We have talked to Prof. David Jones, who have developed both MP-specific tool MEMPACK and generic contact prediction tools PSICOV and MetaPSICOV and informed us that MEMPACK is not as good as MetaPSICOV. A recent paper(Zhang et al., 2016a) proposed a new MP-specific tool COMSAT, compared 12 MP-specific and MP-independent contact prediction tools and showed that MP-specific tools have no significant advantage over the best MP-independent tools. To further verify this, we have tested our deep learning model (trained by non-MPs only) on the 87 membrane proteins tested in the COMSAT paper. Our result shows that our deep model outperforms COMSAT, which was reported to be the best MP-specific contact predictor. See Suppl. Table 2 for detailed results.

**Performance evaluation**

We evaluate our method in terms of both contact prediction and 3D modeling accuracy. We define that a contact is short-, medium- and long-range when the sequence separation of two residues in a contact falls into [6, 11], [12, 23], and ≥24, respectively. We evaluate the accuracy of the top $L/k$ ($k$=10, 5, 2, 1) predicted contacts where L is protein sequence length. The prediction accuracy is defined as the percentage of native contacts among the top $L/k$ predicted contacts. In the case that there are no $L/k$ native contacts in a category, we simply use $L/k$ as the denominator when calculating the accuracy.

We measure the quality of a 3D model by TMscore (Zhang and Skolnick, 2004), which ranges from 0 to 1 indicating the worst and the best quality, respectively. A 3D model with TMscore≥0.6 is likely to have a correct fold while a 3D model with TMscore<0.5 usually does not. TMscore=0.5 is also used by the community as a cutoff to judge if a model has a correct fold or not.

# DATA AND SOFTWARE AVAILABILITY

The web server implementing the deep learning method for contact prediction is publicly available at

http://raptorx.uchicago.edu/ContactMap/. The code and the list of non-membrane training proteins will be provided upon request to the Lead Contact. The list of 510 membrane proteins is available in the Supplemental File. The predicted contacts and 3D models of the 510 membrane proteins are also publicly available at Mendeley Data (https://data.mendeley.com/datasets/4wht7k4knt/1).

## ADDITIONAL RESOURCES

CAMEO: www.cameo3d.org

# Supplementary materials for the following manuscript:

Sheng Wang, Zhen Li, Yizhou Yu and Jinbo Xu. Folding membrane proteins by deep transfer learning.

**Table S1.** A list of 510 non-redundant membrane proteins with solved structures in PDB, related to *Table 1 and STAR Methods section "Data for model parameter optimization and test"*.

| | | | | | | | | | | | | |
|---|---|---|---|---|---|---|---|---|---|---|---|---|
| 1a0sP | 1pw4A | 2evuA | 2lnlA | 2wpvB | 3cn5A | 3kvnA | 3udcA | 4chvA | 4il3A | 4or2A | 4wgvA | 5c8jI |
| 1ar1B | 1q16C | 2f1cX | 2lomA | 2wsc1 | 3cx5C | 3l11A | 3ug9A | 4cskA | 4in5H | 4p6vB | 4wmzA | 5cfbA |
| 1bccE | 1q90A | 2f93B | 2loqA | 2wsc3 | 3d31C | 3lnmB | 3ukmA | 4czbA | 4in5L | 4p6vC | 4x5mA | 5ctgA |
| 1bctA | 1q90B | 2f95B | 2lorA | 2wscF | 3ddlA | 3lw54 | 3um7A | 4d5bA | 4j05A | 4p6vD | 4xk83 | 5d0yA |
| 1bhaA | 1qcrD | 2fynB | 2losA | 2wscG | 3dhwA | 3lw5H | 3uq7A | 4d6tD | 4j72A | 4p6vE | 4xnkA | 5dirA |
| 1c17M | 1qd6C | 2ge4A | 2lotA | 2wscH | 3dinE | 3m71A | 3ux4A | 4d6tG | 4j7cI | 4p6vF | 4xnvA | 5doqA |
| 1e7pC | 1qleC | 2gfpA | 2lp1A | 2wscK | 3dl8C | 3mk7A | 3v2wA | 4d6tJ | 4jkvA | 4p79A | 4xu4A | 5doqB |
| 1ehkB | 1rh5B | 2gr7A | 2m0qA | 2wscL | 3dl8E | 3mk7B | 3v5sA | 4d6uD | 4k1cA | 4pgrA | 4xxjA | 5ee7A |
| 1fftB | 1rh5C | 2gr8A | 2m20A | 2wswA | 3dwoX | 3mk7C | 3vmqA | 4djiA | 4kjrA | 4phzA | 4xydB | 5ek0A |
| 1fftC | 1rwtA | 2h8aA | 2m67A | 2wwbB | 3dwwA | 3mktA | 3vouA | 4dojA | 4knfA | 4pirA | 4y25A | 5ekeA |
| 1fw2A | 1s5lB | 2h8pC | 2m6bA | 2wwbC | 3dzmA | 3mp7A | 3vr8C | 4dveA | 4kppA | 4px7A | 4y28G | 5eulE |
| 1fx8A | 1s5lE | 2hdfA | 2m7gA | 2x4mA | 3effK | 3mp7B | 3vr8D | 4dxwA | 4kt0F | 4q2eA | 4y28K | 5ezmA |
| 1gzmA | 1s5lX | 2ibzG | 2m8rA | 2xq2A | 3eh3A | 3njtA | 3vwiA | 4e1tA | 4kt0K | 4qncA | 4y28L | 5f1cA |
| 1h2sB | 1sqqK | 2ibzI | 2mafA | 2xutA | 3ejzA | 3nymA | 3wdoA | 4ea3A | 4ky0A | 4qndA | 4y7jA | 5fn2B |
| 1h6s1 | 1t16A | 2iubA | 2mfrA | 2y5yA | 3emnX | 3o0rB | 3wmfA | 4ezcA | 4l6rA | 4qtnA | 4ymkA | 5gaeh |
| 1izlA | 1tlwA | 2j58A | 2mgyA | 2y69D | 3emoA | 3o7pA | 3wmm1 | 4f35A | 4l6v6 | 4quvA | 4ymsC | 5gaqA |
| 1izlC | 1tqqA | 2j7aC | 2mm8A | 2y69G | 3fhhA | 3ohnA | 3wmmM | 4f4lA | 4l6v8 | 4r1iA | 4ytpC | 5garO |
| 1jb0K | 1uunA | 2jafA | 2mmuA | 2y69I | 3fidA | 3orgA | 3wo7A | 4fqeA | 4ltoA | 4rdqA | 4ytpD | 5hk1A |
| 1k24A | 1uynX | 2jlnA | 2mn6A | 2y69J | 3g67A | 3oufA | 3wvfA | 4fuvA | 4m58A | 4rfsS | 4z34A | 5i1mV |
| 1kf6C | 1vclA | 2jo1A | 2mpnA | 2y69K | 3gi8C | 3p5nA | 3wxvA | 4g1uA | 4m64A | 4ri2A | 4z3nA | 5i20A |
| 1kf6D | 1vf5B | 2jp3A | 2mxbA | 2y69L | 3hd6A | 3pjsK | 3x29A | 4g7vS | 4mbsA | 4rjwA | 4z7fA | 5i32A |
| 1kqfB | 1vf5D | 2k0lA | 2n4xA | 2y69M | 3hw9A | 3pjzA | 3x2rA | 4g80I | 4meeA | 4rl8A | 4zp0A | 5i6cA |
| 1kqfC | 1wrgA | 2k21A | 2n6lA | 2yevB | 3iyzA | 3pwhA | 3x3bA | 4gbyA | 4mndA | 4rl9A | 4zr0A | 5i6zA |
| 1kzuA | 1xioA | 2k73A | 2n7qA | 2yevC | 3iz1A | 3q7kA | 3ze3A | 4gd3A | 4mqsA | 4rlcA | 4zr1A | 5id3A |
| 1lghA | 1xl4A | 2k9pA | 2nmrA | 2yiuA | 3j08A | 3qe7A | 3zevA | 4gx5A | 4mt4A | 4rngA | 4zw9A | 5iofA |
| 1m56B | 1yc9A | 2kluA | 2nq2A | 2ynkA | 3j1zP | 3qnqA | 3zjzA | 4gycB | 4n74A | 4rp8A | 5a1sA | 5irxA |
| 1m56D | 1yewC | 2kogA | 2nr9A | 2z73A | 3j9tR | 3qraA | 3zk1A | 4h33A | 4n75A | 4ryiA | 5a40A | 5ivaA |
| 1m57A | 1yq3C | 2ks9A | 2nrgA | 2ziyA | 3jbrE | 3rbzA | 3zuxA | 4he8A | 4njnA | 4s0vA | 5a63C | 5iwsA |
| 1mm4A | 1yq3D | 2ksdA | 2o01F | 2zjsE | 3jcuD | 3rgwS | 4a2nB | 4he8C | 4nppA | 4tkrA | 5a63D | 5ixmB |
| 1mprA | 1zrtE | 2kseA | 2oarA | 2zxeB | 3jcuH | 3rkoA | 4atvA | 4he8D | 4ntjA | 4tq3A | 5a6eB | 5jagA |
| 1n7lA | 1zzaA | 2ksfA | 2pnoA | 2zxeG | 3jcuK | 3rkoB | 4aw6A | 4hkrA | 4nykA | 4tquM | 5abbZ | |
| 1nekC | 2a0lA | 2ksrA | 2q67A | 3a2sX | 3jcuR | 3rkoC | 4b4aA | 4hqjE | 4o6mA | 4tquN | 5araT | |
| 1nekD | 2a9hA | 2kyhA | 2q7mA | 3a7kA | 3jcuS | 3rkoD | 4bemJ | 4httA | 4o6yA | 4twkA | 5araW | |
| 1o5wA | 2akhA | 2l35A | 2qomA | 3anzA | 3jcuW | 3rkoF | 4bgnA | 4huqS | 4o9pA | 4u15A | 5awwG | |
| 1occD | 2akhB | 2l8sA | 2r6gF | 3b4rA | 3jcuX | 3rkoG | 4bog3 | 4huqT | 4o9pB | 4u4tA | 5awwY | |
| 1oedC | 2bg9A | 2lckA | 2r6gG | 3b5dA | 3jcuZ | 3s0xA | 4bpmA | 4hw9A | 4o9uB | 4u9lA | 5awzA | |
| 1orsC | 2bl2A | 2lhfA | 2vpwC | 3b9wA | 3jycA | 3sljA | 4bwzA | 4hycA | 4od4A | 4uc1A | 5aymA | |
| 1p49A | 2cpbA | 2lkgA | 2w1pA | 3bryA | 3k3fA | 3sybA | 4c9jA | 4hyoA | 4ogqC | 4us3A | 5azbA | |
| 1p4tA | 2d57A | 2llyA | 2wjqA | 3chxB | 3kj6A | 3tijA | 4cadC | 4hzuS | 4oh3A | 4v1fA | 5bwkE | |
| 1p7bA | 2ervA | 2lmeA | 2wpdJ | 3chxC | 3kp9A | 3tx3A | 4cfgA | 4iffA | 4oo9A | 4wd7A | 5c6oA | |

**Figure S1.** Case study of one CAMEO target 5jkiA, related to section *"Blind test in CAMEO"* and Figure 4. (A) The long- and medium-range contact prediction accuracy of our methods, MetaPSICOV, CCMpred, and EVfold (Web Server). (B-D) The overlap between top L predicted all-range contacts and the native contact map. A grey, red and green dot represents a native contact, a correct prediction and a wrong prediction, respectively. (E) The superimposition between our predicted model (in red) and the native structure (in green).

|  | Long range accuracy | | | | Medium range accuracy | | | |
|---|---|---|---|---|---|---|---|---|
|  | L | L/2 | L/5 | L/10 | L | L/2 | L/5 | L/10 |
| Our method | 0.658 | 0.883 | 1.000 | 1.000 | 0.185 | 0.351 | 0.659 | 0.864 |
| MetaPSICOV | 0.554 | 0.820 | 0.977 | 1.000 | 0.158 | 0.279 | 0.523 | 0.727 |
| CCMpred | 0.495 | 0.703 | 0.773 | 0.818 | 0.131 | 0.207 | 0.477 | 0.682 |
| EVfold(web) | 0.514 | 0.712 | 0.773 | 0.841 | 0.126 | 0.207 | 0.432 | 0.727 |

(A)

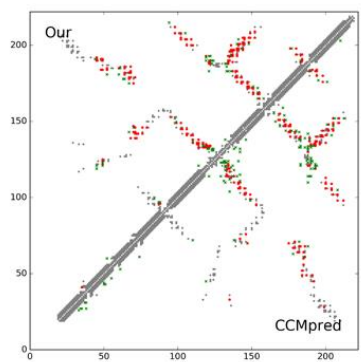

(B)

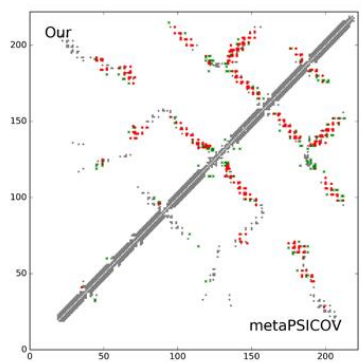

(C)

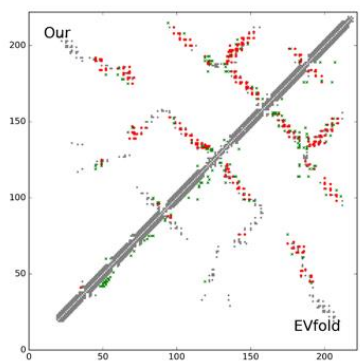

(D)

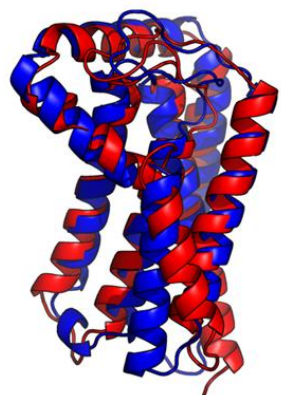

(E)

**Figure S2.** Case study of one CAMEO target 5l0wA, related to section *"Blind test in CAMEO" and Figure 4*. (A) The long- and medium-range contact prediction accuracy of our methods, MetaPSICOV and CCMpred. (B-C) The overlap between top L predicted all-range contacts and the native contact map. A grey, red and green dot represents a native contact, a correct prediction and a wrong prediction, respectively. (D) The superimposition between our predicted model (in red) and the native structure (in green).

|  | Long range accuracy | | | | Medium range accuracy | | | |
| --- | --- | --- | --- | --- | --- | --- | --- | --- |
|  | L | L/2 | L/5 | L/10 | L | L/2 | L/5 | L/10 |
| Our method | 0.397 | 0.674 | 0.889 | 1.000 | 0.103 | 0.207 | 0.444 | 0.778 |
| metaPSICOV | 0.250 | 0.391 | 0.528 | 0.722 | 0.098 | 0.163 | 0.278 | 0.389 |
| CCMpred | 0.087 | 0.109 | 0.222 | 0.333 | 0.016 | 0.033 | 0.056 | 0.056 |

(A)

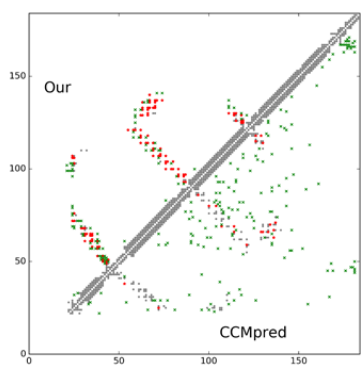

(B)

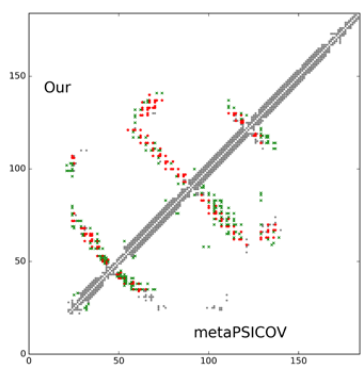

(C)

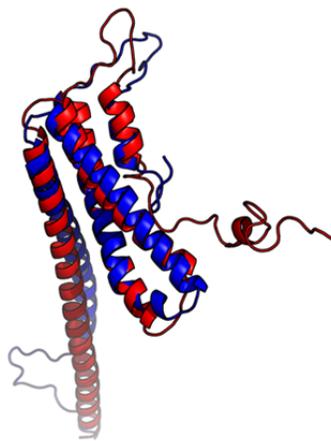

(D)

**Supplementary Figure 3.** Top 20 long-range 5x5 contact occurrence patterns, related to section "*Why does deep transfer learning work?*" and STAR method.

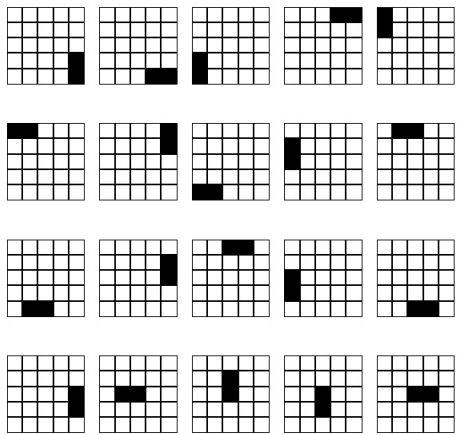

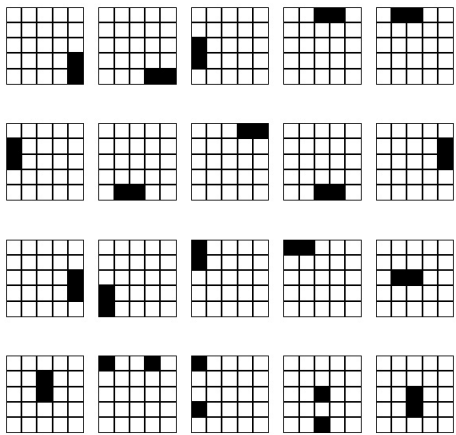

**Table S2.** Contact prediction accuracy on the 87 COMSAT test proteins, *related to STAR Method and Table 1*. Only contacts between two transmembrane segments are evaluated. Acc, Cov, Sp and Mcc represent accuracy, coverage, specificity and Mathew correlation coefficient, respectively. Note that here our result is based upon the deep model trained without using any membrane proteins while COMSAT was trained by some membrane proteins. The result of COMSAT is taken from its paper. Following the COMSAT paper, the results of the latter three methods are calculated on top $L_m$ predicted contacts where $L_m$ is the length of transmembrane regions in a test protein.

| Accuracy of top $L_m$ predicted contacts when a contact is defined by $C_\alpha$-$C_\alpha$ distance less than 14Å. | | | | | | | | | | | | | |
|---|---|---|---|---|---|---|---|---|---|---|---|---|---|
| Method | >=6 | | | | >=12 | | | | >=24 | | | |
| | Acc | Cov | Sp | Mcc | Acc | Cov | Sp | Mcc | Acc | Cov | Sp | Mcc |
| CCMpred | 0.63 | 0.07 | 0.98 | 0.15 | 0.61 | 0.07 | 0.98 | 0.15 | 0.57 | 0.08 | 0.97 | 0.14 |
| MetaPSICOV | 0.73 | 0.08 | 0.99 | 0.19 | 0.72 | 0.08 | 0.99 | 0.19 | 0.69 | 0.10 | 0.98 | 0.19 |
| Our Method | 0.86 | 0.10 | 0.99 | 0.23 | 0.85 | 0.10 | 0.99 | 0.23 | 0.82 | 0.12 | 0.98 | 0.24 |
| COMSAT | 0.65 | 0.05 | 0.99 | 0.11 | 0.63 | 0.054 | 0.99 | 0.11 | 0.61 | 0.052 | 0.99 | 0.10 |
| Accuracy of top $L_m$ predicted contacts when a contact is defined by $C_\beta$-$C_\beta$ distance less than 8Å. | | | | | | | | | | | | | |
| Method | >=6 | | | | >=12 | | | | >=24 | | | |
| | Acc | Cov | Sp | Mcc | Acc | Cov | Sp | Mcc | Acc | Cov | Sp | Mcc |
| CCMpred | 0.32 | 0.25 | 0.98 | 0.26 | 0.31 | 0.26 | 0.97 | 0.26 | 0.29 | 0.27 | 0.96 | 0.25 |
| MetaPSICOV | 0.39 | 0.31 | 0.98 | 0.32 | 0.38 | 0.31 | 0.98 | 0.32 | 0.35 | 0.33 | 0.96 | 0.31 |
| Our Method | 0.59 | 0.46 | 0.98 | 0.48 | 0.58 | 0.46 | 0.98 | 0.48 | 0.53 | 0.49 | 0.97 | 0.47 |
| COMSAT | 0.43 | 0.14 | 0.99 | 0.21 | 0.43 | 0.14 | 0.99 | 0.21 | 0.44 | 0.14 | 0.98 | 0.21 |